\documentclass[12pt]{article}
\usepackage{a4}
\usepackage{epsfig,url}
\usepackage{lineno}

\newcommand\oo{\omega_{12}}
\newcommand\Do{\Delta\omega}
\newcommand\dt{\Delta t}
\newcommand\tR{\tilde R}
\begin{document}
\title{Real-time system identification of superconducting cavities with
  a recursive least-squares algorithm: closed-loop operation}
\author{V. Ziemann, Uppsala University, 75120 Uppsala, Sweden}
\date{August 29, 2023}
\maketitle
\begin{abstract}\noindent
  We simulate a recursive least-squares estimator to determine the bandwidth
  $\omega_{12}$ and the detuning $\Delta\omega$ of a cavity that is controlled
  with a low-level RF feedback system and we present a comprehensive analysis
  of the convergence and asymptotic behavior of the algorithm for static and
  time-varying systems. 
\end{abstract}
%
%
\section{Introduction}
\label{sec:intro}
Superconducting acceleration cavities are used to accelerate protons~\cite{SNS,ESS},
electrons~\cite{CEBAF,XFEL,CBETA}, and heavy ions~\cite{FRIB,SPIRAL2,HELIAC}, both
with pulsed~\cite{SNS,XFEL} and with continuous beams~\cite{CEBAF,JLABFEL}. Owing
to the low losses, the cavities have a very narrow bandwidth on the order of Hz for
bare cavities and a few 100\,Hz for cavities equipped with high-power couplers.
In order to efficiently cool these cavities with liquid helium are the cavities
made of rather thin material, which makes them easily deformable and this changes
their resonance frequency, often by an amount comparable to their bandwidth.
In pulsed operation, the dominant deformation comes from the electro-magnetic
pressure of the field inside the cavity, the Lorentz-force detuning~\cite{LFD},
while cavities operated continuously are perturbed by so-called microphonics~\cite{ANA1},
caused by pressure variations of the liquid helium bath or mechanical perturbations,
for example, by reciprocating pumps or by malfunctioning equipment. As a consequence
of these perturbations, the cavities are detuned and force the power generator to
increase their output to maintain fields necessary for stable operation of the beams.
This reduces the efficiency of the system and requires an, often substantial, overhead
of the power generation, forcing it to operate at a less than optimal working point.
To avoid this sub-optimal mode of operation and  to compensate the detuning, many
accelerators employ active tuning systems that use stepper motors and
piezo-actuators~\cite{TUNER} to squeeze the cavities back in tune, which requires
diagnostic systems to measure the detuning.
These measurements are usually
based on comparing the phase of the signal that excites the cavity, measured with
a directional coupler just upstream of the input coupler, to the phase of the
field inside the cavity, measured by a field probe or antenna inside the cavity.
Both analog~\cite{ANA1,ANA2} and digital~\cite{SCHILCHER,PLAWSKI} signal processing
systems are used; often as part of the low-level radio-frequency (LLRF) feedback
system that stabilizes the fields in the cavity. Even more elaborate systems,
based on various system identification algorithms, were used~\cite{RYBA,CZARSKI}.
All these algorithms normally rely on low-pass filtering the often noisy signals
from the directional couplers and antennas in order to provide a reliable estimate
of the cavity detuning and the bandwidth.
\par
In this report, we focus on a complementary algorithm that exploits correlations
between the {\em changes} of the input signal that the LLRF system prescribes and
the ensuing {\em changes} of the fields, as measured by the field probe. Modern
digital LLRF systems often operate at digitization rates of several Msamples per
second, making a large number of ``change measurements'' available. We subject this
wealth of data to a recursive least-squares (RLS) algorithm to extract the bandwidth  
$\oo$ and the detuning $\Do$ of a cavity as fit parameters. Such algorithms can be
efficiently implemented inside the LLRF and require only moderate numerical
overhead. Moreover, the difference between the continuously improving estimates
of the fit parameters and the ``true'' values---the so-called estimation
error---approaches zero~\cite{LAIWEI}! Finally, this type of algorithm actually
benefits from the measurement noise of the field probe, because it agitates the
feedback and then exploits the ensuing correlations between the input to the
cavity and the field inside. We must, however, point out that the slow convergence
of the algorithm, especially for low-noise systems, makes it more suitable for
continuous-wave operation than for pulsed operation.
\par
In the following sections, we first introduce the model of the cavity and the
feedback and derive a simple proportional controller using optimal control theory
whereas the analysis of PID controllers is deferred to appendices. After
transforming the model to discrete time we develop the RLS algorithm to identify
the cavity parameters in Section~\ref{sec:sysid} and analyze its convergence in
Section~\ref{sec:conv}. In the following two sections we generalize the method
to efficiently deal with time-varying parameters and analyze its convergence.
In Section~\ref{sec:pulse} we briefly address pulsed systems before closing.
\section{Model}
\label{sec:model}
Accelerating cavities can be described by an equivalent circuit composed of a
resistor $R$, an inductance $L$, and a capacitor $C$, all connected in parallel.
This circuit is then excited by a current $I$. In~\cite{VZAPB}
the following differential equation for the envelope of the voltage $V$ in the
cavity is derived 
\begin{equation}\label{eq:dVdt}
  \frac{dV}{dt} =-\left(\frac{\hat\omega}{2Q_L}+\frac{i\hat\omega\delta}{2}\right) V +\frac{\hat\omega R I}{2Q_L(1+\beta)n}
  \quad\mathrm{with}\quad
  \delta=\frac{\omega}{\hat\omega}-\frac{\hat\omega}{\omega}=-\frac{2\Delta\omega}{\hat\omega}\ ,
\end{equation}
where $\omega$ is the frequency of the generator and $\hat\omega^2=1/LC$ is the
resonance frequency of the cavity. The two are usually close, but not necessarily
equal and $\Delta\omega=\hat\omega-\omega$ is the difference between them.
Moreover, $Q_L=Q_0/(1+\beta)$ is the loaded $Q$-value and $Q_0=R\sqrt{C/L}$ is
what's called the unloaded $Q$-value, whereas $\beta$ is the coupling of the
antenna that feeds power into the cavity and $n$ is the winding ratio of
the transformer that normally models the coupler. With the abbreviations
$\hat\omega/2Q_L=\oo$ and $\hat\omega\delta/2=\Delta\omega$, we can rewrite
Equation~\ref{eq:dVdt} in the standard form
\begin{equation}
\frac{dV}{dt} =-(\oo+i\Do )V  +\oo \tR I
\end{equation}
where we defined the abbreviation $\tR=R/(1+\beta)n$ to simplify the notation.
Throughout this report, we assume that all currents and voltages are baseband
signals, that is after the down-mixer and before before the up-mixer.
\par
By splitting the voltage and the current into real and imaginary part, $V=V_r+iV_i$
and $I=I_r+iI_i$ we obtain two equations, one for the real and one for the imaginary
part of the voltage. Assembling the two equations in the form of a matrix leads
us directly to the following state-space representation 
\begin{equation}\label{eq:ss}
  \left(\begin{array}{c} \frac{dV_r}{dt}\\ \frac{dV_i}{dt} \end{array}\right)
  =\left(\begin{array}{cc} -\oo & -\Do \\ \Do& -\oo \end{array}\right)
  \left(\begin{array}{c} V_r \\ V_i \end{array}\right)
  +\left(\begin{array}{cc} \oo\tR & 0 \\ 0& \oo\tR \end{array}\right) 
  \left(\begin{array}{c} I_r \\ I_i \end{array}\right)
\end{equation}
of the system that describes the dynamics of the cavity voltage powered by a
generator that provides the currents.
\par
Normally we want to control a system around some desired voltage level $\hat V_r$ and
$\hat V_i$, which is the set point of the controller. The currents, we assume that they
already include the beam currents, that produce these voltages are easily found by
setting the left-hand side in Equation~\ref{eq:ss} to zero and solving for the currents.
This leads us to
\begin{equation}
  \left(\begin{array}{c} \hat I_r \\ \hat I_i \end{array}\right)
  =\frac{1}{\tR}\left(\begin{array}{cc} 1 & \Do/\oo \\ -\Do/\oo & 1 \end{array}\right)
  \left(\begin{array}{c} \hat V_r \\ \hat V_i \end{array}\right)\ .
\end{equation}
The dynamics around this operating point is then given by expanding the voltages
around the steady state with $V_r=\hat V_r+v_r$ and  $V_i=\hat V_i+v_i$, where
$v_r$ and $v_i$ are small deviations from the desired set point that we want to
stabilize by changing the currents, which we represent likewise by $I_r=\hat I_r+i_r$
and  $I_i=\hat I_i+i_i$. Inserting in Equation~\ref{eq:ss} then leads us to
\begin{equation}\label{eq:sss}
  \left(\begin{array}{c} \frac{dv_r}{dt}\\ \frac{dv_i}{dt} \end{array}\right)
  =\left(\begin{array}{cc} -\oo & -\Do \\ \Do& -\oo \end{array}\right)
  \left(\begin{array}{c} v_r \\ v_i \end{array}\right)
  +\left(\begin{array}{cc} \oo\tR & 0 \\ 0& \oo\tR \end{array}\right) 
  \left(\begin{array}{c} i_r \\ i_i \end{array}\right)\ ,
\end{equation}
which describes essentially the same dynamical system as Equation~\ref{eq:ss},
but now it describes the dynamics around the chosen set point $\hat V_r$ and
$\hat V_i$.
Equation~\ref{eq:sss} is in the standard form of a linear dynamical
system $\dot{\vec v}=\bar A\vec v +\bar B\vec i$ where $\vec v$ is the column
vector of voltages and $\vec i$ to that of the currents. The matrices $\bar A$
and $\bar B$ correspond to those in Equation~\ref{eq:sss} and are given by
\begin{equation}\label{eq:AB}
  \bar A=\left(\begin{array}{cc} -\oo & -\Do \\ \Do& -\oo \end{array}\right)
  \qquad\mathrm{and}\qquad
  \bar B=\left(\begin{array}{cc} \oo\tR & 0 \\ 0& \oo\tR \end{array}\right) \ .
\end{equation}
In order to control this system we use an optimal controller. The discussion of
PID-type regulators is deferred to the appendices.
\section{Optimal control feedback}
\label{sec:oc}
An optimal controller is characterized by minimizing the functional
\begin{equation}\label{eq:oc}
  J[\vec v,\vec i]=\int \left[\vec v^{\top}Q_v\vec v +\vec i^{\top}Q_i\vec i\right]  dt\ ,
\end{equation}
where $Q_v$ and $Q_i$ are weights to specify whether the algorithm should favor
small values of the state $\vec v$ or of the controller $\vec i$. We will
use $Q_v={\mathbf 1}$ and $Q_i=Z^2{\mathbf 1}$, where $Z$ is a constant with
the units of an impedance that makes the units of $\vec v$ (volt) and $\vec i$
(ampere) commensurate in the definition of $J$. Moreover, $\mathbf{1}$ is the
$2\times2$ unit matrix.
\par
The theory to design optimal controllers~\cite{FYF,KIRK} for time-invariant systems
relies on finding the solution $K$ of the stationary Riccati equation
\begin{equation}\label{eq:ric}
  0=-Q_x -\bar A^{\top} K - K\bar A + K\bar BQ_u^{-1}\bar B^{\top}K\ ,
\end{equation}
where $\bar A$ and $\bar B$ are defined in Equation~\ref{eq:AB}.
Since the detuning $\Do$ is normally small and might even oscillate around zero
if it is caused by microphonics~\cite{ANA1}, we will set $\Do$ to zero when determining
$K$. In that case all matrices appearing in Equation~\ref{eq:ric} are diagonal
and the equation separates into two identical scalar equations, which read
\begin{equation}
0=-1+\oo\kappa + \kappa\oo+(\oo^2\tR^2/Z^2)\kappa^2
\end{equation}
for $\kappa$ defined by $K=\kappa\mathbf{1}$. Solving this quadratic equation
results in
\begin{equation}
\kappa=-\frac{Z^2}{\oo\tR^2}\left[1\mp\sqrt{1+\frac{\tR^2}{Z^2}}\right]\ .
\end{equation}
The control law~\cite{FYF} relating $\vec i$ and $\vec v$ is then given by
$\vec i=-Q_u^{-1}\bar B^{\top} K \vec v$ or, in components, we obtain
\begin{equation}\label{eq:claw}
  \left(\begin{array}{c} i_r \\ i_i \end{array}\right)
  =\frac{1}{\tR} \left[1\mp\sqrt{1+\frac{\tR^2}{Z^2}}\right]
 \left(\begin{array}{c} v_r \\ v_i \end{array}\right) \ .
\end{equation}
We see that it depends on the ratio of $\tR$ and $Z$. We can select it to
make the feedback work harder to minimize the voltages (small $Z$) or
to keep the control currents small (large $Z$).
\par
It remains to determine which sign of the root to use.
Inserting the control law from Equation~\ref{eq:claw} into Equation~\ref{eq:sss},
we eliminate the currents $i_r$ and $i_i$ in favor of the voltages $v_r$ and $v_i$
and, after some algebra, arrive at
\begin{equation}
  \left(\begin{array}{c} \frac{dv_r}{dt}\\ \frac{dv_i}{dt} \end{array}\right)=
  \mp\oo \sqrt{1+\frac{\tR^2}{Z^2}}
  \left(\begin{array}{c} v_r \\ v_i \end{array}\right)\ , 
\end{equation}
which makes it obvious that we need to pick the negative sign of the root in order
to make the feedback system stable. The control law is thus given by Equation~\ref{eq:claw}
with the negative sign of the root.
\par
We can summarize the system we will analyze by
\begin{equation}\label{eq:cts}
  \dot{\vec v}=\bar A\vec v + \bar B\vec i + noise
  \qquad\mathrm{and}\qquad
  \vec i=K_p\vec v
\end{equation}
where the feedback gain $K_p=\frac{1}{\tR} \left[1-\sqrt{1+\frac{\tR^2}{Z^2}}\right]$
is given in Equation~\ref{eq:claw}, which turns out to be a proportional controller
that minimizes Equation~\ref{eq:oc}. We also added a noise source to the right-hand
side of the equation in order to account for the process noise in the system.
\section{Simulation}
\label{sec:sim}
For the simulations we will convert the continuous-time system from Equation~\ref{eq:cts}
to discrete time with time step $\Delta t$, which corresponds to the sampling time if the
system is implemented digitally. By replacing the derivatives of the voltages by
finite-difference equations
\begin{equation}
  \frac{d\vec v}{dt}\to\frac{\vec v_{t+1}-\vec v_t}{\Delta t}
\end{equation}
where we label the time steps by $t$. With these substitutions, Equation~\ref{eq:sss}
becomes
\begin{equation}\label{eq:dsa}
  \vec v_{t+1} 
  =A \vec v_t  + B \vec i_t +\vec w_t
  \quad\mathrm{with}\quad A=\left(\begin{array}{cc} 1-\oo \dt & -\Do \dt\\ \Do \dt& 1-\oo \dt \end{array}\right)\ ,
\end{equation}
$B=\oo \dt \tR\mathbf{1}$, and the process noise $\vec w_t$. We assume that the noise is
uncorrelated and has magnitude $\sigma$. It is thus characterized by its expectation
value $E\left\{\vec w_t\vec w_s^{\top}\right\}=\sigma^2\delta_{ts}\mathbf{1}$. The feedback
algorithm looks the same as for the continuous-time system
\begin{equation}\label{eq:dsb}
  \vec i_t  =\frac{1}{\tR} \left[1-\sqrt{1+\frac{\tR^2}{Z^2}}\right] \vec v_t\ .
\end{equation}
Iterating the system of Equations~\ref{eq:dsa} and~\ref{eq:dsb} shows that the
system is indeed stabilized by the feedback algorithm and the rms magnitude
of the voltages compared to that of the currents is indeed given by the ratio
$\tR/Z$ in Equation~\ref{eq:dsb}.
\section{System identification}
\label{sec:sysid}
Now we turn to the task of extracting $\oo \dt$ and $\Do \dt$ from continuously
recording the voltages and currents. In order to isolate the sought parameters
we rewrite Equation~\ref{eq:dsa} in the form
\begin{equation}
  \vec v_{t+1}=\left({\mathbf 1}+F\right)\vec v_t + B\vec i_t
    \qquad\mathrm{with}\qquad
 F=\left(\begin{array}{rr} -\oo \dt & -\Do \dt\\ \Do \dt& -\oo \dt \end{array}\right)\ .  
\end{equation}
and $B=\oo \dt \tR\mathbf{1}$. For the time being we ignore the noise $\vec w_t$
and rewrite this equation as
\begin{equation}\label{eq:yy}
  \vec v_{t+1}-\vec v_t = F \vec v_t + \oo \dt \tR\vec i_t\\ .
\end{equation}
Moreover, we rewrite $F\vec v_t +\oo \dt \tR\vec i_t $ as
\begin{eqnarray}\label{eq:yy2}
  F\vec v_t +\oo\dt \tR\vec i_t &=& -\oo\dt \left(\begin{array}{c} v_r \\ v_i \end{array}\right)_t
  + \Do \dt \left(\begin{array}{r} -v_i \\ v_r \end{array}\right)_t
  +\oo \dt \left(\begin{array}{r} \tR i_r \\ \tR i_i \end{array}\right)_t
  \nonumber\\
  &=& \left(\begin{array}{rr} -v_r+\tR i_r & -v_i  \\ -v_i+\tR i_i & v_r \end{array}\right)_t
  \left(\begin{array}{c} \oo \dt \\ \Do \dt \end{array}\right)\ .
\end{eqnarray}
We now introduce the abbreviations
\begin{equation}\label{eq:defGy}
  G_t = \left(\begin{array}{rr} -v_r+\tR i_r & -v_i  \\ -v_i+\tR i_i & v_r \end{array}\right)_t
  \qquad\mathrm{and}\qquad
  \vec y_{t+1}= \vec v_{t+1}-\vec v_t
\end{equation}
and stack Equation~\ref{eq:yy} for consecutive times on top of each other.
In this way, we obtain a vastly overdetermined system of equations to determine
$\oo\dt$ and $\Do \dt$
\begin{equation}\label{eq:UT}
  \left(\begin{array}{c} \vec y_2 \\ \vec y_3 \\ \vdots \\ \vec y_{T+1}\end{array}\right)
  = U_T  \left(\begin{array}{c} \oo \dt \\ \Do \dt \end{array}\right)
  \quad\mathrm{with}\quad
  U_T=\left(\begin{array}{c} G_1 \\ G_2 \\ \vdots \\ G_T\end{array}\right)
\end{equation}
that we solve in the least-squares sense with the Moore-Penrose pseudo-inverse
\begin{equation}\label{eq:ls}
  \vec q_T=\left(\begin{array}{c} \oo \dt \\ \Do \dt \end{array}\right)_T=
  \left( U_T^{\top} U_T\right)^{-1} U_T^{\top}
  \left(\begin{array}{c} \vec y_2 \\ \vec y_3 \\ \vdots \\ \vec y_{T+1}\end{array}\right) \ .
\end{equation}
Here we introduce the abbreviation $\vec q_T$ to denote the estimated parameters
at time step $T$.
\par
It turns out that we can avoid lengthy evaluations by calculating Equation~\ref{eq:ls}
recursively. With the definition $P_T^{-1}=U^{\top}_TU_T$, its initial value
$P_0=p_0{\mathbf 1}$, and the definition of $U_T$ from Equation~\ref{eq:UT} we
express $P_{T+1}$ through $P_T$ in the following way
\begin{eqnarray}\label{eq:PTT}
  P_{T+1}^{-1}&=&U^{\top}_{T+1}U_{T+1}\nonumber\\
              &=&p_0{\mathbf 1}+G_1^{\top}G_1+G_2^{\top}G_2+\dots+G_T^{\top}G_T+G_{T+1}^{\top}G_{T+1}\\
              &=& P_T^{-1}+G_{T+1}^{\top}G_{T+1}\ . \nonumber
\end{eqnarray}
It turns out that retrieving $P_{T+1}$ from inverting $P_T^{-1}+G_{T+1}^{\top}G_{T+1}$ can be
efficiently done with the Woodbury matrix identity~\cite{NR}
\begin{equation}
  (A+UV^{\top})^{-1}=A^{-1}- A^{-1}U(\mathbf{1}+V^{\top}A^{-1}U)^{-1}V^{\top} A^{-1}\ .
\end{equation}
With the substitutions $A^{-1}=P_T$, $U=G^{\top}_{T+1}$, and $V^{\top}=G_{T+1}$  we obtain
\begin{eqnarray}\label{eq:upp}
  P_{T+1}&=&P_T-P_TG_{T+1}^{\top}\left(\mathbf{1}+G_{T+1}P_TG_{T+1}^{\top}\right)^{-1} G_{T+1}P_T\nonumber\\
  &=& \left[\mathbf{1}-P_TG_{T+1}^{\top}\left(\mathbf{1}+G_{T+1}P_TG_{T+1}^{\top}\right)^{-1} G_{T+1}\right]P_T\ ,
\end{eqnarray}
where the second representation is occasionally convenient to use.
\par
\begin{figure}[tb]
\begin{center}
\includegraphics[width=0.47\textwidth]{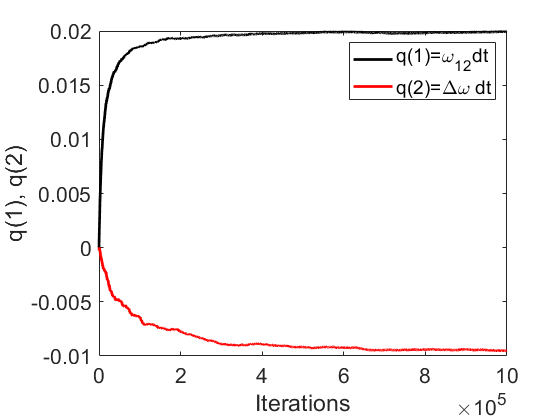}
\includegraphics[width=0.47\textwidth]{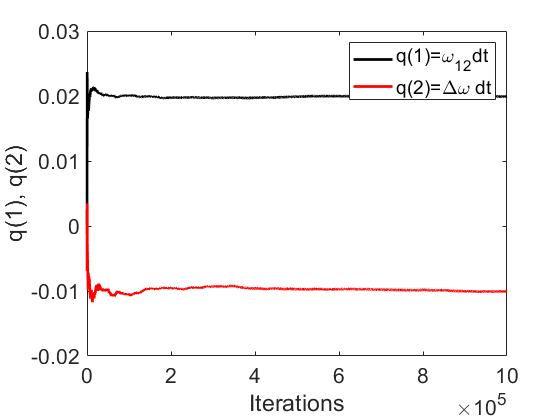}
\end{center}
\caption{\label{fig:conv}The convergence of the fit parameters $\vec q=(\oo \dt,\Do \dt)$ as a
  function of the iterations for $\sigma=10^{-3}$ (left) and $\sigma=10^{-2}$ (right). We
  parameterize the feedback by $Z/\tR=0.5$ which makes $K_p=-1.24/\tR$.}
\end{figure}
We now turn to finding $\vec q_{T+1}$ by writing Equation~\ref{eq:ls} for $T+1$
\begin{eqnarray}\label{eq:upq}
  \vec q_{T+1}&=& P_{T+1} \left(G_1^{\top}\vec y_2+G_2^{\top}\vec y_3+\dots
                  + G_T^{\top}\vec y_{T+1} +G_{T+1}^{\top}\vec y_{T+2}\right)\nonumber\\
              &=&\left[\mathbf{1}-P_TG_{T+1}^{\top}\left(\mathbf{1}+G_{T+1}P_TG_{T+1}^{\top}\right)^{-1} G_{T+1}\right]P_T\\
                  &&\qquad\times
                  \left(\sum_{t=1}^T G_t^{\top}\vec y_{t+1} +G_{T+1}^{\top}\vec y_{T+2}\right)\nonumber\\
              &=&\left[\mathbf{1}-P_TG_{T+1}^{\top}\left(\mathbf{1}+G_{T+1}P_TG_{T+1}^{\top}\right)^{-1} G_{T+1}\right]
                  \left( \vec q_T+P_tG_{T+1}^{\top}\vec y_{T+2}\right)\ . \nonumber
\end{eqnarray}
We point out that only the $2\times 2$ matrix $\mathbf{1}+G_{T+1}P_TG_{T+1}^{\top}$ has to be
inverted in every time step. Equation~\ref{eq:upp} and~\ref{eq:upq} constitute the algorithm
to continuously update estimates for the two components of $\vec q$, the bandwidth
$q(1)=\oo \dt$ and the detuning $q(2)=\Do \dt$, as new voltage measurements $\vec v_{T+1}$
and currents $\vec i_{T+1}$---both enter in $G_{T+1}$ and $\vec y_{T+2}$---become available.
We refer to the code on github~\cite{GITHUB} for the details of the implementation.
\par
We also point out that the structural similarity between Equation~\ref{eq:ss} and~\ref{eq:sss}
implies that the algorithm works both for small changes around a set point and for large
excitations that occur in pulsed operation of cavities that we briefly address in
Section~\ref{sec:pulse}. The algorithm encapsulated in Equations~\ref{eq:upp} and~\ref{eq:upq}
just efficiently extracts the cavity parameters from the reaction of the output voltages
to excitations of the cavity. This works by either deliberately exciting the cavity in an
open-loop configuration or when it is controlled by a feedback system in a closed-loop
configuration. We will focus on the latter in most of this report, because this mode
of operation allows the system to operate parasitically during ``production''  runs of
the accelerator. Moreover, it permits a detailed analysis of the convergence of the
algorithm.
\par
Figure~\ref{fig:conv} illustrates the performance of the algorithm to determine the ``true''
values $\vec q_0=(0.02,-0.01)$ over $10^6$ iterations with the feedback system operating
with a proportional controller, parameterized by $Z/\tR=0.5$, which makes $K_p=-1.24/\tR$.
The numerical values of $\vec q_0$ are chosen to be rather large (cavity bandwidth of
3\,kHz in a 352\,MHz cavity, sampled at 1\,Msamples/s)
in order to visualize the performance of the algorithm. The two plots show the convergence
for a noise level  of $\sigma=10^{-3}$ on the left and $\sigma=10^{-2}$ on the right.
We see that the convergence is much slower, though also smoother, for the lower noise level.
This is no surprise, because the noise causes excursions in the voltages that the feedback
system tries to compensate. More noise causes larger excursions for both the voltages and
the currents and from correlating these values, the system parameters are extracted with the
help of Equation~\ref{eq:ls} and its recursive versions Equation~\ref{eq:upp} and~\ref{eq:upq}.
\par
Figure~\ref{fig:conv} motivates a number of questions about the convergence of the algorithm, such
as the relevant parameters, apart from the noise level $\sigma$, that determine the rate of convergence.
Moreover, the asymptotically reachable precision is an important quantity to determine. 
\section{Convergence}
\label{sec:conv}
In this section we use the methods, developed in~\cite{ZZ1}, to analyze a recursive
least-squares algorithm, its rate of convergence and its asymptotic behavior. We immediately
note that $\left(U_T^{\top}U_T\right)^{-1}\sigma^2=P_T\sigma^2$ is the empirical covariance
matrix that contains the squares of error bars~\cite{FYF} of fit parameters $\vec q_T$
on its diagonal.
The evolution of $P_T$ can be analyzed with the help of Equation~\ref{eq:upp}. From numerical
experiments we know that $P_T$ decreases quite rapidly such that, in this analysis,  we
approximate $\mathbf{1}+G_{T+1}P_TG_{T+1}^{\top}$ by $\mathbf{1}$ which simplifies
Equation~\ref{eq:upp} to
\begin{equation}\label{eq:Ptconv}
  P_{T+1}\approx P_T-P_TG_{T+1}^{\top}G_{T+1}P_T\ .
\end{equation}
Here $G_t$ depends on the voltages $\vec v_t$ and currents $\vec i_t$ as defined in
Equation~\ref{eq:defGy}. For the controller from Equation~\ref{eq:cts} the currents
are related to the voltages via $\vec i_t=K_p\vec v_t$, such that we can write the
matrix $G_t$ in the form
\begin{equation}
  G_t = \left(\begin{array}{rr}
                (-1+\tR K_p) v_r & -v_i \\  (-1+\tR K_p)v_i & v_r
              \end{array}\right)_t
            = \vec v_t^{\top} H = (v_r, v_i)_t H 
\end{equation}
where $H$ is a $2\times 2 \times 2$ array that is defined as follows
\begin{equation}\label{eq:defH}
  H(1,:,:)=\left(\begin{array}{cr}  -1+\tR K_p & 0 \\ 0 & 1\end{array}\right)
  \quad\mathrm{and}\quad
  H(2,:,:)=\left(\begin{array}{cr}  0 & -1 \\ -1+\tR K_p & 0\end{array}\right)\ .
\end{equation}
Here we use MATLAB~\cite{MATLAB} notation with colons denoting the second
and third argument. In the same way, we can write the transpose $G_t^{\top}$ as
\begin{equation}
G_t^{\top} = \left(\begin{array}{cc}
                (-1+\tR K_p) v_r & (-1+\tR K_p)v_i\\ -v_i & v_r
              \end{array}\right)_t
            = \bar H \vec v_t 
\end{equation}
with the $2\times 2 \times 2$ array $\bar H$ defined by
\begin{equation}\label{eq:defbH}
  \bar H(:,:,1) =\left(\begin{array}{cr}  -1+\tR K_p & 0 \\ 0 & 1\end{array}\right)
  \quad\mathrm{and}\quad
  \bar H(:,:,2)=\left(\begin{array}{rc}  0 & -1+\tR K_p\\ -1 & 0\end{array}\right)\ .
\end{equation}
With these definitions we express $G_{T+1}^{\top}G_{T+1}$ as
\begin{equation}
  G_{T+1}^{\top}G_{T+1} = \bar H \vec v_{T+1}\vec v_{T+1}^{\top} H\ .
\end{equation}
In order to remove the dependence on the process noise $\vec w_t$ we replace
$G_{T+1}^{\top}G_{T+1}$ by its expectation value $E\left\{ G_{T+1}^{\top}G_{T+1} \right\}
=\bar H E\left\{\vec v_{T+1}\vec v_{T+1}^{\top}\right\} H$ and for this we need to
calculate $\vec v_{T+1}$ in terms of all previous $\vec w_t$ and then perform
the expectation value.
\par
We obtain $\vec v_{T+1}$ by repeatedly inserting Equation~\ref{eq:dsb} into Equation~\ref{eq:dsa},
which leads to
\begin{equation}\label{eq:evvt}
\vec v_{T+1} = (A+ K_p B)^{T+1} \vec v_0 + \sum_{s=0}^{T} (A+ K_p B)^s\vec w_{T-s}\ .
\end{equation}
If the largest eigenvalue of $A+\kappa B$ is smaller than unity, the initial value $\vec v_0$
``dies out'' after a while and we can neglect this term and obtain for
$E\left\{\vec v_{T+1}\vec v_{T+1}^{\top}\right\}$
\begin{eqnarray}\label{eq:Evv}
 E\left\{\vec v_{T+1}\vec v_{T+1}^{\top}\right\}
  &=& E\left\{\sum_{s=0}^{T}\sum_{r=0}^{T} (A+K_p B)^s\vec w_{T-s}\vec w_{T-r}^{\top}
      \left((A+K_p B)^{\top}\right)^r\right\}\nonumber\\
  &=& \sigma^2 \sum_{s=0}^{T} \left( (A+K_p B)(A+ K_p B)^{\top}\right)^s + o(1)\ ,
\end{eqnarray}
where we used the statistics of the noise defined immediately after Equation~\ref{eq:dsa}.
Here $o(1)$ denotes a quantity that vanishes in the limit of large $T$. Noting that the
product of matrices in the sum is symmetric, we can diagonalize it and find
\begin{equation}
  (A+K_p B)(A+K_p B)^{\top} = O \Lambda O^{\top} = \lambda\mathbf{1}\ ,
\end{equation}
where $O=\mathbf{1}$, because the matrix on the left-hand side is anti-symmetric and,
additionally, has equal entries on the diagonal. Therefore also the two diagonal elements
of $\Lambda$ are equal and we denote them by $\lambda$. Now inserting this expression into
Equation~\ref{eq:Evv}, we obtain
\begin{equation}\label{eq:Evv2}
  E\left\{\vec v_{T+1}\vec v_{T+1}^{\top}\right\}
  = \sigma^2 \left(\sum_{s=0}^{T} \lambda^s\right)\mathbf{1}
  \approx \frac{\sigma^2}{1-\lambda}\mathbf{1} \ ,
\end{equation}
where we use that the sum over powers of $\lambda$ approximates a geometric series
of which we used the asymptotic value.
We point out that the right-hand side of Equation~\ref{eq:Evv2} no longer depends
on $T$ and is therefore constant, such that the expectation value
$Q=E\left\{ G_{T+1}^{\top}G_{T+1} \right\}$ is given by
\begin{equation}\label{eq:EGG}
  Q=E\left\{ G_{T+1}^{\top}G_{T+1} \right\} = \frac{\sigma^2}{1-\lambda} \bar H H
  =\frac{\sigma^2}{1-\lambda}\left(\begin{array}{cc} 2(-1+\tR K_p)^2 & 0\\ 0 & 2\end{array}\right)\ .
\end{equation}
Here we we evaluate the product of $\bar H H$ by denoting  the matrix elements
of $\bar H$ by $\bar H_{i\alpha\beta}$, those of $H$ by $H_{\beta\alpha j}$, and then 
summing over the indices $\alpha$ and $\beta$ in the following fashion
\begin{equation}
(\bar H H)_{ij} = \sum_{\alpha=1}^2 \sum_{\beta=1}^2 \bar H_{i\alpha\beta} H_{\beta\alpha j}\ .
\end{equation}
Writing down the four parts of the double sum and assembling the appropriate
matrix elements from Equation~\ref{eq:defH} and~\ref{eq:defbH} directly leads
to the result stated in Equation~\ref{eq:EGG}.
\par
For convenience we introduce the abbreviations
\begin{equation}\label{eq:defmu}
  \mu_1= \frac{2(-1+\tR K_p)^2}{1-\lambda}
  \qquad\mathrm{and}\qquad
  \mu_2=\frac{2}{1-\lambda}
\end{equation}
such that the diagonal elements of $Q$ are $\sigma^2\mu_1$ and $\sigma^2\mu_2$.
\par
\begin{figure}[tb]
\begin{center}
  \includegraphics[width=0.8\textwidth]{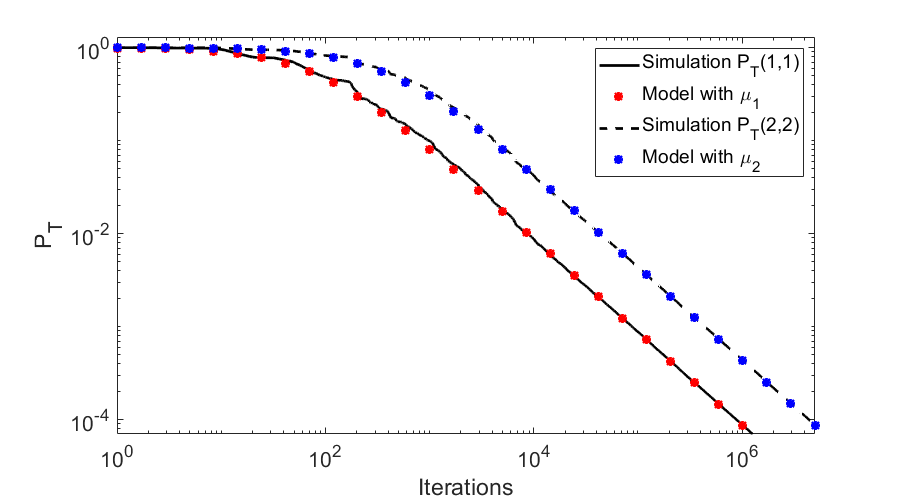}
\end{center}
\caption{\label{fig:pt}The evolution of $P_T(1,1)$ (solid) and $P_T(2,2)$ (dashed) for
  $4\times 10^6$ iterations from a simulation and evaluated from Equation~\ref{eq:pta}
  with $\mu_1$ (red asterisks) and with $\mu_2$ (blue asterisks). The used parameters
  correspond to those from the right-hand image in Figure~\ref{fig:conv}.}
\end{figure}
We  now return to equation~\ref{eq:Ptconv} and  replace $G_{T+1}^{\top}G_{T+1}$ by its
asymptotic expectation value $Q$. We also note that both $Q$ and the initial value
of $P_0=p_0\mathbf{1}$ are diagonal. In other words the two degrees of freedom
do not mix and we can consider them independently. In the following paragraphs
$p_T$ denotes one of the diagonal elements of $P_T$ and $\sigma^2\mu$ is the
corresponding diagonal element in $Q$. The temporal evolution of $p_T$ is
then given by
\begin{equation}\label{eq:approxpt}
  p_{T+1} \approx p_T -\sigma^2\mu p_T^2
\end{equation}
that we solve in a smooth approximation as
\begin{equation}
  \frac{p_{T+1}-p_T}{dT}  =-\sigma^2\mu p_T^2
  \qquad\mathrm{or}\qquad
  \frac{dp_T}{dT} = -\sigma^2\mu p_T^2
\end{equation}
where $dT=T+1-T=1$. This equation has the solution
\begin{equation}\label{eq:pta}
  p_T = \frac{p_0}{1+\sigma^2\mu p_0T}\ .
\end{equation}
We find that $p_T$ shrinks proportional to $T_s/T$, where the time scale $T_s$
is given by $T_s=1/\mu\sigma^2p_0$ and only depends on the noise level $\sigma^2$ and
on $\mu$, which is defined in Equation~\ref{eq:defmu} and encapsulates the essential
dynamics of the system. 
\par
Figure~\ref{fig:pt} shows $p_T$ from a simulation with parameters corresponding to
those used in the right-hand image in Figure~\ref{fig:conv}. Here the black solid
line shows $P_T(1,1)$ from the simulation and the red asterisks the values produced by
Equation~\ref{eq:pta}  with $\mu=mu_1$. The dashed black line shows  $P_T(2,2)$ with
the model superimposed as blue asterisks. We find good agreement for both curves over
the full range of iterations.
\par
\begin{figure}[tb]
\begin{center}
  \includegraphics[width=0.8\textwidth]{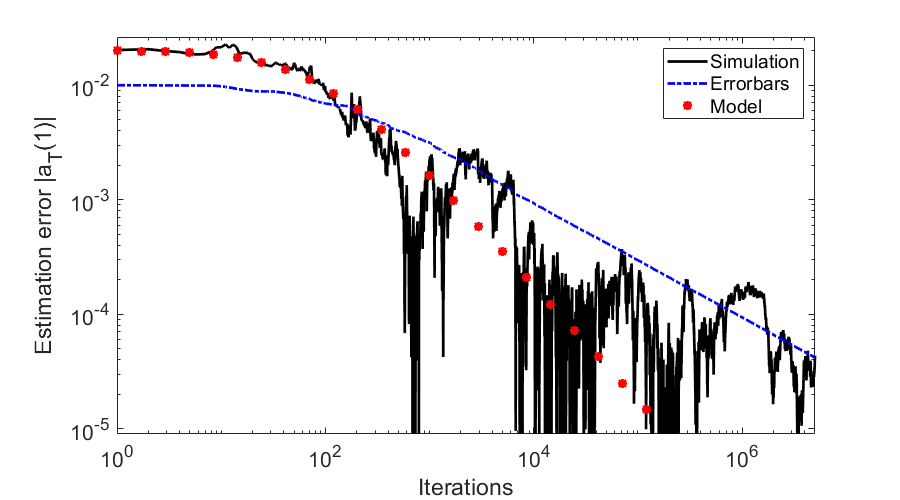}
\end{center}
\caption{\label{fig:at}The evolution of the estimation error $|a_T(1)|=|q_T(1)-q_0(1)|$
  (black), the model from Equation~\ref{eq:ata} (red asterisks), and the error bars
  $\sqrt{P_T(1,1)\sigma^2}$ (blue dots).}
\end{figure}
Let us now introduce the estimation error $\vec a_T=\vec q_T-\vec  q_0$ where we denote the
``true'' parameters by $\vec q_0 = (\oo \dt, \Do \dt)$ and use Equation~\ref{eq:upq} to
calculate its time evolution. As before we approximate $\mathbf{1}+G_{T+1}P_TG_{T+1}^{\top}$
by $\mathbf{1}$ and find
\begin{equation}
\vec q_{T+1}=\left[\mathbf{1}-P_TG_{T+1}^{\top}G_{T+1}\right]\left( \vec q_T+P_tG_{T+1}^{\top}\vec y_{T+2}\right)\ . 
\end{equation}
Moreover, by combining Equations~\ref{eq:yy} and~\ref{eq:yy2} we see that $\vec y_{T+2}$ can be
written as
\begin{eqnarray}
  \vec y_{T+2}=\vec v_{T+2}-\vec v_{T+1} = G_{T+1}\vec q_0\ .
\end{eqnarray}
Replacing $\vec y_{T+2}$ in Equation~\ref{eq:upq} leads to
\begin{eqnarray}
  \vec q_{T+1}
  &=&\left[\mathbf{1}-P_TG_{T+1}^{\top}G_{T+1}\right]\left( \vec q_T+P_tG_{T+1}^{\top}G_{T+1}\vec q_0\right)
      \nonumber\\
  &\approx&  \vec q_T - P_TG_{T+1}^{\top}G_{T+1}\vec q_T +P_tG_{T+1}^{\top}G_{T+1}\vec q_0\ ,
\end{eqnarray}
where we omitted the higher-order contribution with the product of two $P_T$. Subtracting $\vec q_0$
on both sides and introducing the estimation error $\vec a_T=\vec q_T -\vec q_0$ allows us to write
\begin{equation}
  \vec a_{T+1} = \vec a_T +  P_TG_{T+1}^{\top}G_{T+1}\vec a_T \ .
\end{equation}
As previously, we now approximate  $G_{T+1}^{\top}G_{T+1}$ by its expectation value $Q$ from
Equation~\ref{eq:EGG} and obtain
\begin{equation}
  \vec a_{T+1} \approx \vec a_T +  P_TQ\vec a_T 
\end{equation}
Since both $P_T$ and $Q$ are diagonal the system separates into two independent equations
that differ only by their respective value of the parameter $\mu$ from Equation~\ref{eq:defmu}.
We therefore omit the arrow on the estimation error $a_T$; it is either one component or the
other that has to obey the equation
\begin{equation}\label{eq:aaa}
  a_{T+1} = a_T - p_T \sigma^2\mu a_T\ .
\end{equation}
As before, we transform this equation into its smooth approximation and obtain
\begin{equation}
\frac{da_T}{dT}=-\sigma^2\mu p_T  a_T = -\sigma^2\mu\frac{p_0}{1+\sigma^2\mu p_0T} a_T\ ,
\end{equation}
where we replaced $p_T$ by its approximation from Equation~\ref{eq:pta}. Integrating
this equation is straightforward and leads to the solution
\begin{equation}\label{eq:ata}
  a_T = \frac{a_0}{1+p_0\sigma^2\mu T}\ ,
\end{equation}
where $a_0$ is the difference between the initial estimate of the corresponding component
of $\vec q$ and the ``true'' value. Equation~\ref{eq:ata} then describes the temporal
evolution of the difference between the parameter estimate corresponding component of
$ \vec q_T$ and that of the ``true'' value $\vec q_0$. 
\par
Figure~\ref{fig:at} shows the evolution of $a_T$ for the first parameter $q(1)$ from the
same simulation used to generate Figure~\ref{fig:pt} as solid black line. The values
from the model in Equation~\ref{eq:ata} are shown as red asterisks. We observe that
the two curves agree reasonably well up to about $100$ iterations, where the simulation
starts to significantly deviate from the model. On the same figure, we show the error bars
of the fit parameters, approximately given by $\sqrt{P_t(1,1)\sigma^2}$, as the blue
dashed line. Once the estimation errors have the same order of magnitude as the error
bars, replacing the $G_{T+1}^{\top}G_{T+1}$ by its expectation value $Q$ appears to be
no longer a valid approximation. Instead of following Equation~\ref{eq:ata}, the magnitude
of the estimation error follows the magnitude of the error bars, which scale as
$1/\sqrt{T}$. This behavior is also consistent with the very general analysis of the
asymptotic behavior of least-squares algorithms from~\cite{LAIWEI}. Remarkably, the
error bars and with it the estimation errors will asymptotically approach zero, such
that the ``true'' system parameters are determined exactly. This is a consequence that
new information in the form of new measurements is added to improve the estimate
$\vec q_T$ until it reaches the ``true'' parameters. 
\par
As long as $a_T$ is larger than the error bars, the approximation from Equation~\ref{eq:ata}
works reasonably well. We estimate the number of iterations $N_t$ where the transition
from the initial convergence to the asymptotic regime occurs by equating the estimation
error and the error bars given by $\sqrt{p_t\sigma^2}=\sqrt{p_0\sigma^2/(1+p_0\sigma^2\mu)}$.
We find the following value for $N_t$
\begin{equation}
  N_t= \frac{a_0^2-p_0\sigma^2}{\mu p_0^2\sigma^4}\ ,
\end{equation}
where we see that $N_t$ depends on the initial excess of the estimation error above the
noise level $a_0^2- p_0\sigma^2$ and is inversely proportional to the noise level on the
measurement system.
\par
So far, we assumed that the system parameters $\oo \dt$ and $\Do\dt$ are constant, which
ensures that the ``true'' system parameters are asymptotically approached. In most
systems, however, this assumption is not always valid, especially when the cavity is
affected by microphonics. We therefore adapt the algorithm in the
next sections to follow time-varying system parameters.
\section{Time-varying parameters}
\label{sec:tvp}
%
\begin{figure}[tb]
\begin{center}
\includegraphics[width=0.47\textwidth]{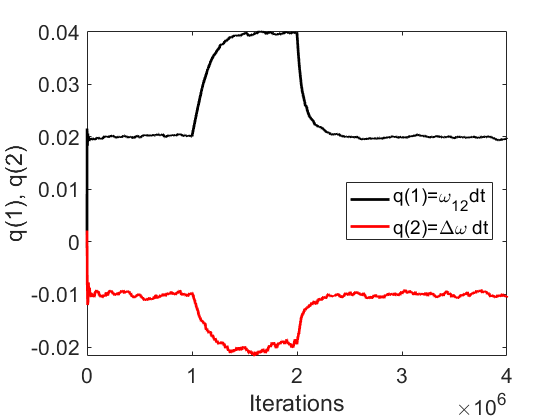}
\includegraphics[width=0.47\textwidth]{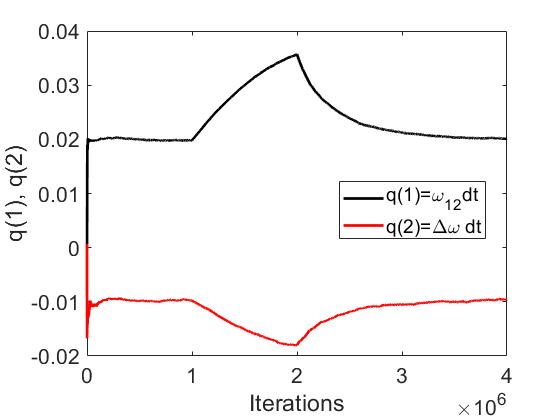}
\end{center}
\caption{\label{fig:Nf}Forgetting parameter $N_f=10^5$ (left) and $N_f=5\times 10^5$ (right).}
\end{figure}
In order to emphasize newly added information we follow~\cite{AW,OP} and introduce
a ``forgetting factor'' $\alpha=1-1/N_f$ where $N_f$ is the time horizon over
which old information is downgraded in the last equality of Equation~\ref{eq:PTT},
which now reads
\begin{equation}
P_{T+1}^{-1}= \alpha P_T^{-1}+G_{T+1}^{\top}G_{T+1}\ . \nonumber
\end{equation}
We see that we only have to replace $P_T$ by $P_T/\alpha$ in the derivation of
Equations~\ref{eq:upp} and~\ref{eq:upq} and find for the update of $P_T$
\begin{equation}\label{eq:uppt}
  P_{T+1}= \frac{1}{\alpha}\left[\mathbf{1}-P_TG_{T+1}^{\top}\left(\alpha+G_{T+1}P_TG_{T+1}^{\top}\right)^{-1} G_{T+1}\right]P_T
\end{equation}
and for the update of the estimated parameters $\vec q_T$
\begin{equation}\label{eq:upqt}
  \vec q_{T+1}=\left[\mathbf{1}-P_TG_{T+1}^{\top}\left(\alpha+G_{T+1}P_TG_{T+1}^{\top}\right)^{-1} G_{T+1}\right]
                  \left( \vec q_T+\frac{1}{\alpha}P_tG_{T+1}^{\top}\vec y_{T+2}\right)
\end{equation}
that are capable of following time-dependent system parameters.
\par
In Figure~\ref{fig:Nf} we show the system parameters $q(1)$ and $q(2)$ over $4\times 10^6$ iterations.
All other parameters correspond to those used in the previous examples. Between iterations $10^6$
and $2\times 10^6$ the values of both parameters are doubled. In the plot on the left-hand side
we use a forgetting time scale $N_f=10^5$ and see that the algorithm tracks the changes,
but the two traces are noisy. For the plot on the right-hand side, $N_f$ is increased to
$5\times 10^5$, which results in a much slower response to follow the changed parameters, albeit
with much less noisy traces.
\par
\begin{figure}[tb]
\begin{center}
  \includegraphics[width=0.47\textwidth]{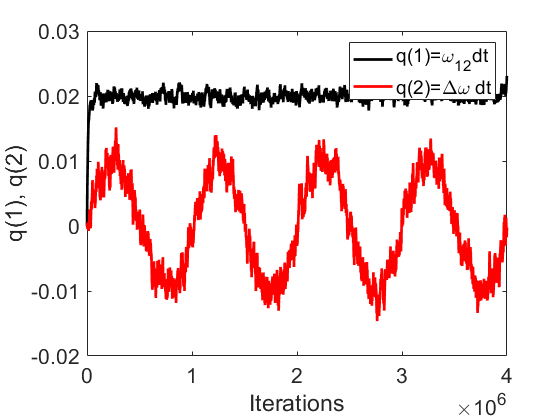}
  \includegraphics[width=0.47\textwidth]{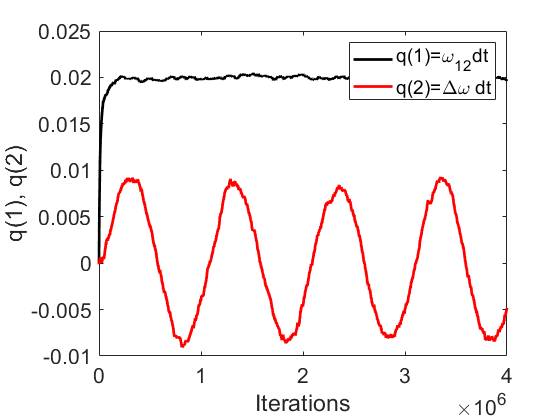}
  \includegraphics[width=0.47\textwidth]{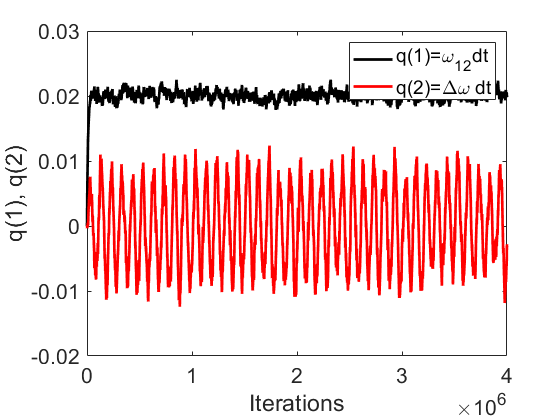}
  \includegraphics[width=0.47\textwidth]{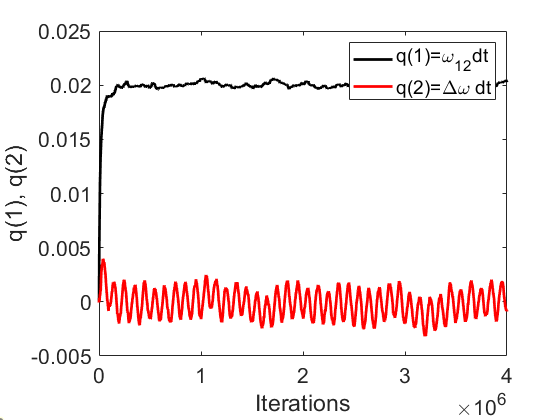}
\end{center}
\caption{\label{fig:osc}Four (top row) and 40 (bottom row) oscillations
  of $\Delta\omega dt$ with $N_f=10^4$ (left column) and $N_f=10^5$
  (right column) for $\sigma=10^{-3}$.} 
\end{figure}
In a second simulation, we keep the $q_0(1)=\oo\dt$ fixed but let $q_0(2)=\Do\dt$ oscillate
around zero with amplitude 0.01, a situation that might be caused by microphonics.
The top row in Figure~\ref{fig:osc} shows the tracked system parameters over
$4\times 10^6$ iterations. When sampling with 1\,Msamples/s the four oscillations
shown correspond to an oscillation of $\Do\dt$ with 1\,Hz. On the top left we use
$N_f=10^4$ and find that the algorithm nicely tracks the oscillations, but the
reconstructed parameters are rather noisy. Increasing $N_f$ to $10^5$ leads to a
rather faithful, and much less noisy, reconstruction of the parameters. The bottom
row in Figure~\ref{fig:osc} shows the corresponding plots for a ten times higher
oscillation period, now corresponding to 10\,Hz. We observe that on the left-hand
side with $N_f=10^4$ the oscillations are clearly recovered, even the amplitude is
approximately correct. The right-hand plot with $N_f=10^5$ shows the amplitude
approximately reduced tenfold. This is a consequence of using $N_f$ with a magnitude
comparable to the oscillation period. Essentially, the oscillations are averaged out.
We conclude that we can recover oscillations that have a period a few times longer
than the ``forgetting'' horizon $N_f$.
\par
That with smaller values of $N_f$ the recovered parameters become more noisy warrants
an analysis of the tradeoff between tracking fast parameter changes and accurate
determination of the parameters.
\section{Convergence revisited}
\label{sec:convt}
%
\begin{figure}[tb]
\begin{center}
  \includegraphics[width=0.47\textwidth]{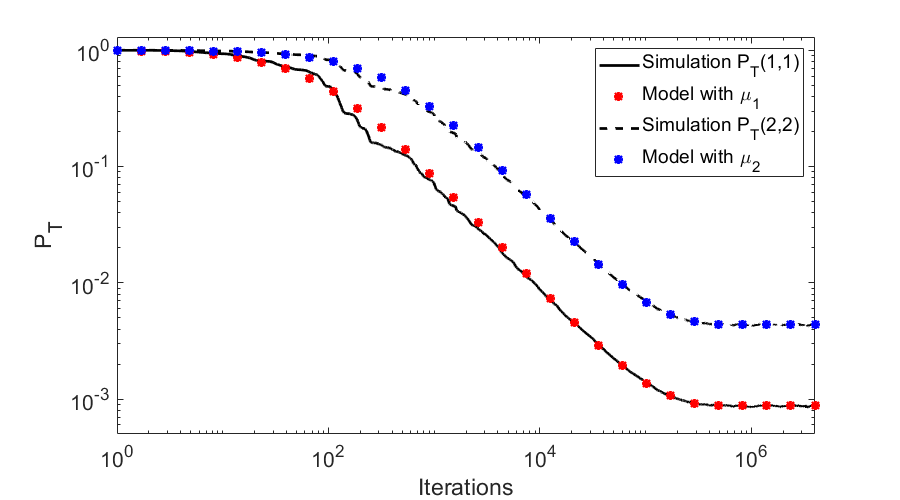}
  \includegraphics[width=0.47\textwidth]{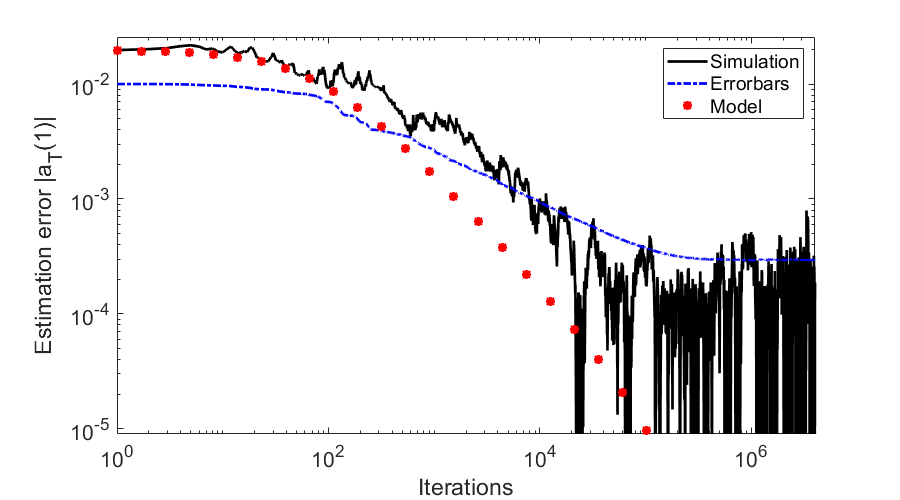}
\end{center}
\caption{\label{fig:asym}The function $p_T$ (left) for $\mu_1$ (red) and $\mu_2$
  (blue) and estimation error (right) from the simulation (black), from
  Equation~\ref{eq:attt} (red asterisks) and the error bars (blue dots) with
  $N_f=10^5$ and $\sigma=0.01$.}
\end{figure}
We adapt the analysis from Section~\ref{sec:conv} to include $\alpha=1-1/N_F$.
All of the calculations and approximation up to Equation~\ref{eq:approxpt} proceed
in the same fashion and its equivalent becomes
\begin{equation}
p_{T+1} \approx \frac{1}{\alpha}\left( p_T-\frac{\mu\sigma^2p_T^2}{\alpha}\right)\ ,
\end{equation}
where we recall that $p_T$ is one of the diagonal entries in $P_T$ and $\mu$ is
the corresponding eigenvalue of $Q$ from Equation~\ref{eq:EGG}. With
$p_{T+1}-p_T \to dp_T/dT$ we rewrite this equation in the smooth approximation as
\begin{equation}
  \frac{dp_T}{dT} = \left(\frac{1-\alpha}{\alpha}\right) p_T - \frac{\mu\sigma^2}{\alpha^2} p_T^2
\end{equation}
which has the solution
\begin{equation}\label{eq:ptb}
  p_T=\frac{p_0}{\mu\sigma^2p_0/\beta\alpha^2-(\mu\sigma^2p_0/\beta\alpha^2-1)e^{-\beta T}}
  \quad\mathrm{with}\quad
  \beta=\frac{1-\alpha}{\alpha}\ .
\end{equation}
The left-hand plot in Figure~\ref{fig:asym} shows good agreement of $p_T$ for the
first diagonal element of $P_T$ from a simulation (solid black line) and from
Equation~\ref{eq:ptb} with $\mu=\mu_1$ (red asterisks). Likewise the temporal
evolution of the second diagonal element (dashed black line) agrees well with
Equation~\ref{eq:ptb} with $\mu=\mu_2$ (blue asterisks). Importantly, we find
that $p_T$ reaches a limiting value $p_{\infty}$ for a large number of iterations.
It is given in the limit $T\to \infty$ of Equation~\ref{eq:ptb}
\begin{equation}
p_{\infty} =\frac{\alpha^2\beta}{\mu\sigma^2} \approx\frac{1}{N_f\mu\sigma^2}
\end{equation}
and favors large values of $N_f$. Clearly, the asymptotic error bars $\sigma(q_i)$
for the fit parameters $q_i$ are given by
\begin{equation}\label{eq:limit}
\sigma(q_i)\approx\sqrt{p_{\infty}\sigma^2} \approx \frac{1}{\sqrt{N_f\mu}}
\end{equation}
no longer approach zero with increasing number of iterations, as they did in
Section~\ref{sec:conv}, but remain finite. Remarkably, this asymptotic resolution
is only determined by $N_f$ and by the parameter $\mu$ from Equation~\ref{eq:defmu}. 
\par
In a similar fashion, replacing the expectation value of $G^{\top}_{T+1} G_{T+1}$ by $Q$,
and performing the same approximations that were used to derive Equation~\ref{eq:aaa} in
Section~\ref{sec:conv} we find that the temporal evolution of each of the estimation
errors is governed by
\begin{equation}
  a_{T+1} = a_T - \frac{\sigma^2\mu}{\alpha} p_T a_T\ .
\end{equation}
Inserting $p_T$ from Equation~\ref{eq:ptb} and replacing the difference $a_{T+1}- a_T$
by the differential $d a_T/dT$, we obtain
\begin{equation}
  \frac{d a_T}{dT} = -\frac{\alpha\beta}{1-ce^{-\beta T}} a_T
  \qquad\mathrm{with}\qquad
  c=1-\frac{\beta\alpha^2}{\mu\sigma^2p_0}\ .
\end{equation}
Separating variables then leads to
\begin{equation}
  \frac{da_T}{a_T} = -\frac{\alpha\beta dT}{1-ce^{-\beta T}}\ .
\end{equation}
With the substitution $z=1-ce^{-\beta T}$ we integrate both sides and find
\begin{equation}
\ln(a/a_0) = \alpha\int\frac{dz}{z(z-1)}=\alpha \ln\left.\left(\frac{z-1}{z}\right)\right\vert_0^T\ .
\end{equation}
Finally, after substituting back $T$ for $z$, we obtain the temporal evolution of $a_T$ 
\begin{equation}\label{eq:attt}
a_T=a_0\ln\left(\frac{(1-c)e^{-\beta T}}{1-ce^{-\beta T}}\right)^{\alpha}\ ,
\end{equation}
where the two components of $\vec a_T$ differ in the value of $\mu$ that enters the
parameter $c$ in Equation~\ref{eq:attt}.
\par
The right-hand plot in Figure~\ref{fig:asym} shows the evolution of the estimation
error of $q(1)=\oo \dt$ for $10^6$ iterations as a solid black line. The model estimate from
Equation~\ref{eq:attt}, shown by the red asterisks, nicely follows the black line
until it crosses the blue line that  indicates the error bars of the estimate.
This is the same type of behavior we already found in Section~\ref{sec:conv}.
%
\section{Pulsed operation}
\label{sec:pulse}
%
\begin{figure}[tb]
\begin{center}
\includegraphics[width=0.47\textwidth]{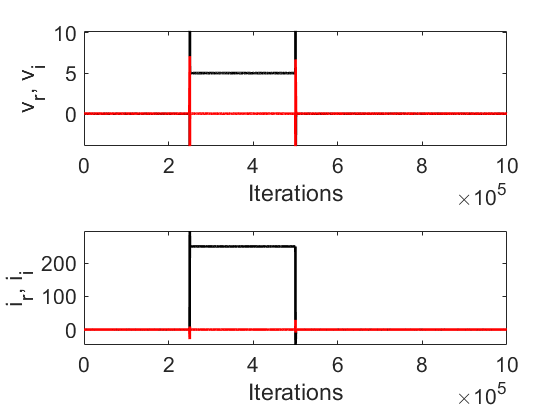}
\includegraphics[width=0.47\textwidth]{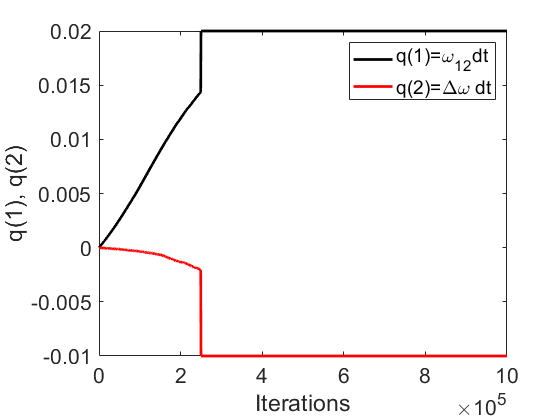}
\end{center}
\caption{\label{fig:pu}The voltages and currents (left) and the
  identified system parameters (right) in pulsed operation.}
\end{figure}
Finally, we tested the performance of the system in pulsed mode, even though it
was conceived for continuously operating systems. The right-hand plots in
Figure~\ref{fig:pu} show the voltages and currents where the set point of the
$v_r$ was increased after $2.5\times 10^5$ iterations and reduced to zero after
$5\times 10^5$ iterations while operating with a proportional-integral controller,
which caused some overshoot and ringing at the rising and falling edges of the
pulse since no effort was made to optimize the feedback parameters. In this
simulation we reduced the noise level to $\sigma=10^{-4}$ such that the convergence
to determine the system parameters, shown on the right-hand plot in
Figure~\ref{fig:pu}, is very slow until, at the start of the pulse, the correct
values are immediately found. The algorithm efficiently exploits the large excursions
of the currents and the voltages during the start of the pulse, determines the
system parameters and rapidly converges to the ``real'' values. This may prove
useful to keep an eye on pulse-to-pulse changes of $q(1)=\oo\dt=\hat\omega \dt/2Q_L$
and thus on the deterioration of $Q_L$ before a quench.
\section{Conclusion and Outlook}
%
\begin{figure}[tb]
\begin{center}
\includegraphics[width=0.47\textwidth]{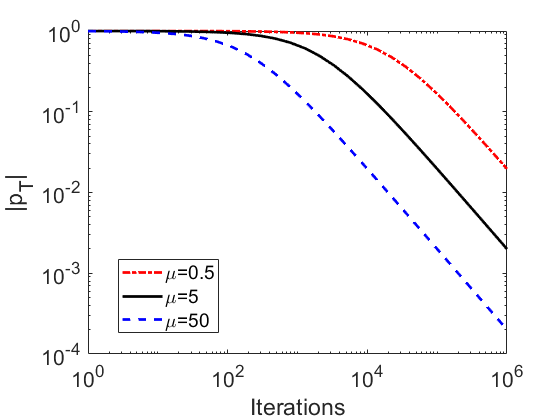}
\includegraphics[width=0.47\textwidth]{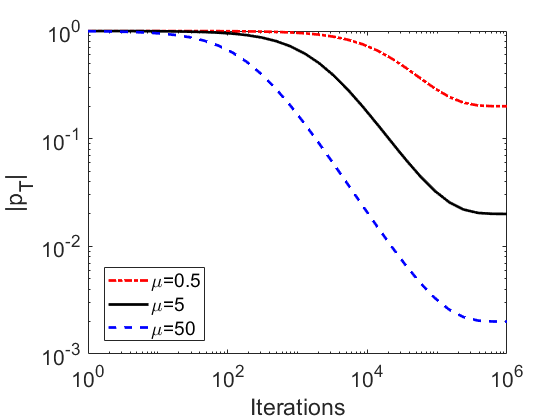}
\end{center}
\caption{\label{fig:cmpmu} The parameters $\vert p_T\vert$ for infinite $N_f$
  from Equation~\ref{eq:pta} (left) and for $N_f=10^5$ (right). In each
  case, three values of $\mu=0.5$ (red), $5$ (black) and $50$ (blue) are
  shown, where larger values of $\mu$ clearly cause faster convergence. On
  the right-hand side the limiting values for large number of iterations
  are clearly seen. In all plots we use $\sigma=0.01$.}
\end{figure}
We worked out an algorithm to determine the cavity bandwidth $\oo$ and the detuning
$\Do$ from LLRF feedback data in real time. The calculations are very efficient and
given by Equations~\ref{eq:upp} and~\ref{eq:upq} for static parameters and by
Equations~\ref{eq:uppt} and~\ref{eq:upqt} for time-varying parameters. The latter
case showed a distinct tradeoff between the ability to follow fast changes and the
achievable resolution, whose limit is given by Equation~\ref{eq:limit}. 
\par
The evolution of the diagonal elements of $P_T$, which determines the data-driven
covariance matrix and thus the error bars, can be analytically described for 
both cases. Its dynamics is entirely determined by the expectation value 
$Q=E\left\{G^{\top}_{T+1}G_{T+1}\right\}$ and its eigenvalues $\mu_1$ and $\mu_2$.
We determine them for different regulators (optimal, PID) in the text and in 
the appendices. The left-hand plot in Figure~\ref{fig:cmpmu} shows the 
evolution of one diagonal element of the covariance matrix $P_T$ for three 
values of $\mu$ where the forgetting horizon $N_f=\infty$ causes it
to decrease without limit. On the other hand, with $N_f=10^5$ iterations
the diagonal elements of $P_T$ saturate at finite values, and thereby finite
error bars, for a large number of iterations. 
\par
We found that the evolution of the estimation error $a_T$ can also be
analytically estimated as long as it is larger than the error bars. In this
regime it follows a $1/T$ dependence. Once it becomes comparable to the error
bars it still decreases, but with a $1/\sqrt{T}$ dependence, which is 
consistent with the general behavior expected~\cite{LAIWEI} for least-squares 
algorithms.
\par
Despite being developed for continuously operated system, the algorithm also
works for pulsed systems and determines the system parameters during
times of change, such as the rise of the pulse. This stimulates the idea to
deliberately introduce small perturbations---so-called dithering---to improve
the convergence of the system identification~\cite{ZZ1}, albeit at the cost of
slightly deteriorating the performance of the feedback system. The algorithm
from Section~\ref{sec:sysid} and~\ref{sec:tvp} robustly filters out any useful
information to improve the estimate of the fit parameters. One might even say
that any process noise is good noise for system identification.
\subsection*{Acknowledgments}
Discussions with Tor Lofnes, Uppsala University are gratefully acknowledged.
%
%
\bibliographystyle{plain}

%
\appendix
\section{Convergence with a PD controller}
\label{sec:appA}
In Section~\ref{sec:conv} we found that the convergence of the algorithm is governed by
$Q$, the expectation value of $E\left\{G^{\top}_{T+1}G_{T+1}\right\}$ and the parameters
$\mu_1$ and $\mu_2$ from Equation~\ref{eq:defmu}. We therefore calculate these
parameters for a proprtional-differential (PD) controller in this appendix for which the
currents $i_t$ are given by
\begin{equation}\label{eq:iPD}
  \vec i_t=K_p\vec v_t -K_d(\vec v_t-\vec v_{t-1})
\end{equation}
instead of Equation~\ref{eq:dsb}. Here $K_p$ is the proportional gain defined near
the end of Section~\ref{sec:oc}. Inserting this expression into Equation~\ref{eq:dsa}
gives us
\begin{equation}
  \vec v_{t+1} = \left(A+B K_p-B K_d\right)\vec v_t + B K_d \vec v_{t-1}\ .
\end{equation}
Introducing the abbreviations $C=A+B K_p-B K_d$ and $D= B K_d$ this defines a
recursion relation that resembles the one for the Fibonacci sequence. Since linear
difference equations are solved by power laws, we use the Ansatz
$\vec v_{t+1}=Z^t\vec w_1$ and find that the matrix $Z$ has to satisfy
\begin{equation}
  Z^2-C Z-D=0
\end{equation}
which is solved by
\begin{equation}
Z_{1,2} = \frac{1}{2}\left( C \pm \sqrt{C^2+4D}\right) \ ,
\end{equation}
where the square root of the matrix $C^2+4D$ can be evaluated in Matlab with the
function {\tt sqrtm()}. Taking the square root can return a matrix with imaginary
entries, which forces us to use the hermitian conjugate (transpose and complex conjugate)
of a matrix $M$, denoted by $M^{\ast}$, instead of the transpose. For matrices with
real entries the hermitian conjugate reverts to the transpose.
\par
With $Z_{1}$ and $Z_2$  known the voltages $\vec v_t$ are given by
$\vec v_{t+1}=\left(c_1 Z_1^t + c_2 Z_2^t\right) \vec w_1$
with constants $c_1$ and $c_2$ that are determined from the initial conditions
$\vec v_1=0$ and $\vec v_2=\vec w_1$ and lead to
\begin{equation}
  0=\left(c_1 Z_1^0 + c_2 Z_2^0\right) \vec w_1
  \qquad\mathrm{and}\qquad
  \vec w_1= \left(c_1 Z_1^1 + c_2 Z_2^1\right) \vec w_1\ .
\end{equation}
Solving for the constants gives us $c_1=-c_2=\left(Z_1-Z_2\right)^{-1}$ such that
\begin{equation}
\vec v_{t+1}=\left(Z_1-Z_2\right)^{-1}\left(Z_1^t - Z_2^t\right) \vec w_1
\end{equation}
which allows us to calculate the influence of the noise $\vec w_1$ at $t=1$ on all
subsequent voltages $\vec v_t$. Conversely, adding the contributions from all previous
noise sources from $t=1$ to $t=T$, we find that $\vec v_{T+1}$ is given by
\begin{equation}
\vec v_{T+1} = \sum_{s=0}^{T} \left(Z_1-Z_2\right)^{-1}\left(Z_1^{s} - Z_2^{s}\right) \vec w_{T+1-s}\ .
\end{equation}
With the voltages known, we assemble $G_{T+1}$ by inserting Equation~\ref{eq:iPD} into 
Equation~\ref{eq:defGy} and find
\begin{eqnarray}\label{eq:defGPD}
  G_t &=& \left(\begin{array}{rr} (-1+\tR(K_p-K_d))v_r & -v_i \\
                  (-1+\tR(K_p-K_d))v_i& v_r\end{array}\right)_t
       +\tR K_d\left(\begin{array}{lc} v_r & 0 \\ v_i & 0\end{array}\right)_{t-1}\nonumber\\
      &=& \vec v_t^{\top} H + \vec v^{\top}_{t-1}J
\end{eqnarray}
with $H$ being defined analogously to Equation~\ref{eq:defH} with the substitution $K_p \to K_p-K_d$
and $J$ is defined by
\begin{equation}\label{eq:defJPD}
  J(1,:,:)=\tR K_d\left(\begin{array}{cr}  1 & 0 \\ 0 & 0\end{array}\right)
  \quad\mathrm{and}\quad
  J(2,:,:)=\tR K_d\left(\begin{array}{cr}  0 & 0 \\ 1 & 0\end{array}\right)\ .
\end{equation}
In the same fashion we define $\bar H$ analogously to Equation~\ref{eq:defbH} and $\bar J$ by
\begin{equation}\label{eq:defbJPD}
  \bar J(:,:,1)=\tR K_d\left(\begin{array}{cr}  1 & 0 \\ 0 & 0\end{array}\right)
  \quad\mathrm{and}\quad
  \bar J(:,:,2)=\tR K_d\left(\begin{array}{cr}  0 & 1 \\ 0 & 0\end{array}\right)\ .
\end{equation}
which allows us to write
\begin{equation}
G^{\top}_t=\bar H \vec v_t +\bar J\vec v_{t-1}
\end{equation}
With these definitions we can express $G^{\top}_{T+1}G_{T+1}$ by
\begin{eqnarray}\label{eq:GGPD}
  G^{\top}_{T+1}G_{T+1}
  &=&\left( \bar H \vec v_{T+1} +\bar J\vec v_T\right)
            \left( \vec v_{T+1}^{\top} H + \vec v_{T}^{\top} J\right)\\
  &=& \bar H \vec v_{T+1}\vec v_{T+1}^{\top} H+  \bar H \vec v_{T+1}\vec v_{T}^{\top} J
     +\bar J\vec v_T\vec v_{T+1}^{\top} H+\bar J\vec v_T\vec v_{T}^{\top} J\nonumber
\end{eqnarray}
and observe that, in order to determine $Q$ we need to calculate the expectation values 
of $\vec v_{T+1}\vec v_{T+1}^{\top}$, $\vec v_{T+1}\vec v_{T}^{\top}$, $\vec v_{T}\vec v_{T+1}^{\top}$, and
$\vec v_{T}\vec v_{T}^{\top}$. Let us therefore consider the cross term $E\left\{\vec v_{T+1}\vec v_{T}^{\top}\right\}$
as an example. The calculations for the other expectation values work the same way
\begin{eqnarray}\label{eq:EvvPDapp}
  E\left\{\vec v_{T+1}\vec v_{T}^{\top}\right\}
  &=& E\left\{\sum_{s=0}^{T}\sum_{r=0}^{T-1}\left(Z_1-Z_2\right)^{-1}\left(Z_1^{s} - Z_2^{s}\right)\right.\nonumber \\
  && \times\left. \vec w_{T+1-s}\vec w_{T-r}^{\top}\left(Z_1^{\ast r}-Z_2^{\ast r}\right)\left(Z_1^{\ast}-Z_2^{\ast}\right)^{-1} \right\}
  \nonumber\\
  &=& \sigma^2\sum_{r=0}^{T-1}\left(Z_1-Z_2\right)^{-1}\left(Z_1^{r+1} - Z_2^{r+1}\right)
      \left(Z_1^{\ast r}-Z_2^{\ast r}\right)\left(Z_1^{\ast}-Z_2^{\ast}\right)^{-1}\nonumber\\
  &=& \sigma^2 \sum_{r=0}^{T-1}\left(Z_1-Z_2\right)^{-1}
      \left[ Z_1\left(Z_1Z_1^{\ast}\right)^r - Z_1\left(Z_1Z_2^{\ast}\right)^r\right.\\
  && \quad\qquad\qquad\left. - Z_2\left(Z_2Z_1^{\ast}\right)^r + Z_2\left(Z_2Z_2^{\ast}\right)^r
     \right]\left(Z_1^{\ast}-Z_2^{\ast}\right)^{-1}\nonumber\\
  &\approx& \sigma^2 \left(Z_1-Z_2\right)^{-1}\left[Z_1\left(\mathbf{1}-Z_1Z_1^{\ast}\right)^{-1}
      -Z_1\left(\mathbf{1}-Z_1Z_2^{\ast}\right)^{-1} \right.\nonumber\\
  && \qquad\left. -Z_2\left(\mathbf{1}-Z_2Z_1^{\ast}\right)^{-1}+Z_2\left(\mathbf{1}-Z_2Z_2^{\ast}\right)^{-1}\right]
     \left(Z_1^{\ast}-Z_2^{\ast}\right)^{-1}\ .
      \nonumber 
\end{eqnarray}
In the second equality we use $E\{\vec w_{T+1-s}\vec w_{T-r}^{\top}\}=\sigma^2\delta_{s(r+1)}\mathbf{1}$ 
and in the last approximate equality we extend the sum to infinity and use the summation formula
for infinite geometric sums that only converges for $\vert Z_{1,2}\vert <1$, which defines the limit 
of stability for the system. With the abbreviations $A_{ij}=\left(\mathbf{1}-Z_iZ_j^{\ast}\right)^{-1}$
and $S=\left(Z_1-Z_2\right)^{-1}$ we write the other expectation values as
\begin{eqnarray}
  E\left\{\vec v_{T+1}\vec v_{T+1}^{\top}\right\}
  &=& S\left[Z_1A_{11}Z_1^{\ast}-Z_1A_{12}Z_2^{\ast}-Z_2A_{21}Z_1^{\ast}+Z_2A_{22}Z_2^{\ast}\right] S^{\ast}\nonumber\\
  E\left\{\vec v_{T}\vec v_{T+1}^{\ast}\right\} 
  &=& S\left[A_{11}Z_1^{\ast}-A_{12}Z_2^{\ast}-A_{21}Z_1^{\ast}+A_{22}Z_2^{\ast}\right] S^{\ast}\\
  E\left\{\vec v_{T}\vec v_{T}^{\ast}\right\} 
  &=& S\left[A_{11}-A_{12}-A_{21}+A_{22}\right] S^{\ast}\ .\nonumber
\end{eqnarray}
In Equation~\ref{eq:GGPD} these expectation values are sandwiched between the arrays $H$ and $J$.
Those terms are evaluated by calculating, for example
\begin{equation}
(\bar H E\left\{\vec v_{T+1}\vec v_{T}^{\top}\right\} J)_{ij} = \sum_{\alpha=1}^2 \sum_{\beta=1}^2\sum_{\gamma=1}^2
\bar H_{i\alpha\beta}M_{\beta\gamma}J_{\gamma\alpha j}
\end{equation}
with the $2\times 2$ matrix $M=E\left\{\vec v_{T+1}\vec v_{T}^{\top}\right\}$. Adding the four terms 
from the second line in Equation~\ref{eq:GGPD} then gives us $Q$, whose eigenvalues are 
$\mu_1$ and $\mu_2$. We refer to the Matlab code from~\cite{GITHUB} for the details of the calculations.
\par
\begin{figure}[tb]
\begin{center}
\includegraphics[width=0.47\textwidth]{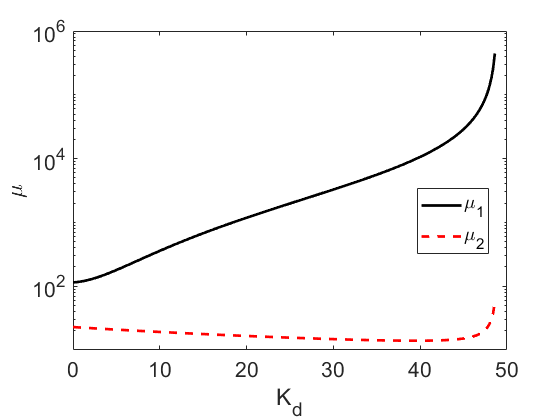}
\includegraphics[width=0.47\textwidth]{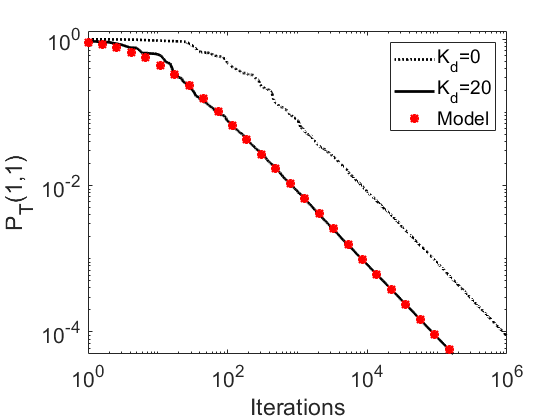}
\end{center}
\caption{\label{fig:Kd}The parameters $\mu$ from Equation~\ref{eq:Evv2} as a function
  of $K_d$ (left) and $P_T(1,1)$ for $K_d=0$ (dotted) and $K_d=20$ (solid) along
  with the model (red asterisks).}
\end{figure}
For the parameters used elsewhere in this report, the left-hand plot in Figure~\ref{fig:Kd} 
shows $\mu_1$ and $\mu_2$ as a function of $K_d$ from zero up to the limit of stability. 
We observe that $\mu_1$ significantly increases, which actually improves the convergence
to determine $q(1)=\oo \dt$, albeit at the expense to determine $q(2)=\Do\dt$.
The right-hand plot shows the evolution
of $P_T(1,1)$ for $K_d=0$ (dotted) and $K_d=20$ (solid black) where $\mu_1=1183$.
We see that with the larger value of $\mu$ the convergence sets in earlier. The
red asterisks are calculated from Equation~\ref{eq:pta} with $\mu=1183$ and show
good agreement with the values from the simulation. The dotted line from a 
simulaton with $K_d=0$ indeed shows slower convergence.
\section{Convergence with a PI controller}
\label{sec:appB}
In this appendix we calculate $\mu_1$ and $\mu_2$ for a proportional-integral (PI) controller, 
whose control law is defined by
\begin{equation}\label{eq:cpi}
  \vec i_t=K_p\vec v_t -K_i\sum_{i=0}^t\vec v_i\ .
\end{equation}
As for the PD controller, inserting Equation~\ref{eq:cpi} into Equation~\ref{eq:dsa}
leads to
\begin{equation}
  \vec v_{t+1} = \left(A+B K_p\right)\vec v_t - B K_i\sum_{i=0}^t\vec v_i\ .
\end{equation}
In order to remove the dependence on the sum, we calculate the difference of voltages
on two consecutive time steps $t+1$ and $t+2$. After some algebra we obtain
\begin{equation}
\vec v_{t+2}-\vec v_{t+1}= \left( A +BK_p-BK_i\right) \vec v_{t+1} -  \left( A +BK_p\right) \vec v_{t}\ .
\end{equation}
Adding $\vec v_{t+1}$ on both sides then leads to
\begin{equation}
\vec v_{t+2}= \left(\mathbf{1}+ A +BK_p-BK_i\right) \vec v_{t+1} - \left( A +BK_p\right) \vec v_{t}\ .
\end{equation}
Introducing the abbreviations $\hat C= \mathbf{1}+A +BK_p-BK_i$ and $\hat D=A+BK_p$ and using the
Ansatz $\vec v_{t+1}=Z^t\vec w_1$ the charactersitic equation for this system becomes
\begin{equation}
  Z^2-\hat CZ+\hat D=0
\end{equation}
which has the solutions
\begin{equation}
  Z_{1,2} = \frac{1}{2}\left( \hat C \pm \sqrt{\hat C^2-4\hat D}\right) 
\end{equation}
such that we can write the general solution as
\begin{equation}
 \vec y_{t+1} = \left( c_1 Z_1^t +c_2 Z_2^t\right)\vec w_1\ .
\end{equation}
As in the previous appendix, we obtain the constants $c_1$ and $c_2$ from matching the
initial values and, as before, find $c_1=-c_2=\left(Z_1-Z_2\right)^{-1}$.
As a consequence of calulating the difference of $\vec v_{t+2}-\vec v_{t+1}$ the
variable $\vec y_t$ is the sum of all $\vec v_i$ and we have to take the difference between
consecutive values in order to obtain 
\begin{equation}
  \vec v_{t+1}=\vec y_{t+1}-\vec y_t = \left(Z_1-Z_2\right)^{-1}
  \left[(Z_1-\mathbf{1})Z_1^{t-1}-(Z_2-\mathbf{1})Z_2^{t-1}\right]\vec w_1\ .
\end{equation}
As for the PD controller, we assemble the matrix $G_t$ from the voltages by inserting the 
control law from Equation~\ref{eq:cpi} into the definition of $G_t$ from Equation~\ref{eq:defGy} 
and obtain
\begin{eqnarray}\label{eq:defGPI}
  G_t &=& \left(\begin{array}{rr} (-1+\tR K_p)v_r & -v_i \\
                  (-1+\tR K_p)v_i& v_r\end{array}\right)_t
       -\tR K_i\left(\begin{array}{lc} y_r & 0 \\ y_i & 0\end{array}\right)_{t}\nonumber\\
      &=& \vec v_t^{\top} H - \vec y^{\top}_{t}K
\end{eqnarray}
where $\vec y_t=\sum_{s=0}^t \vec v_s$ and the array $K$ is defined analogously to $J$ in
Equation~\ref{eq:defJPD} in Appendix~\ref{sec:appA}
\begin{equation}\label{eq:defKPI}
  K(1,:,:)=\tR K_i\left(\begin{array}{cr}  1 & 0 \\ 0 & 0\end{array}\right)
  \quad\mathrm{and}\quad
  K(2,:,:)=\tR K_i\left(\begin{array}{cr}  0 & 0 \\ 1 & 0\end{array}\right)
\end{equation}
and  $\bar K$ analogously to Equation~\ref{eq:defbJPD}. With these definitions
$G^{\top}_{T+1}G_{T+1}$ is given by
\begin{equation}\label{eq:GGPI}
  G^{\top}_{T+1}G_{T+1}
  = \bar H \vec v_{T+1}\vec v_{T+1}^{\top} H - \bar H \vec v_{T+1}\vec y_{T+1}^{\top} J
      - \bar J\vec y_{T+1}\vec v_{T+1}^{\top} H+\bar J\vec y_{T+1}\vec y_{T+1}^{\top} J\ .
\end{equation}
Calculating the expectation values is somewhat lengthy but follows the same
spirit as in the previous appendix and we refer to the code on~\cite{GITHUB} for the
details. As an example, the expectation value of $\vec v_{T+1}\vec y_{T+1}^{\top}$ reads
\begin{eqnarray}
  E\{\vec v_{T+1} \vec y_{T+1}^{\top}\}
  &\approx& \sigma^2\left[ R_1\left(\mathbf{1}-Z_1Z_1^{\ast}\right)^{-1}Z_1^{\ast}
            -R_1\left(\mathbf{1}-Z_1Z_2^{\ast}\right)^{-1}Z_2^{\ast}\right.\\
  &&      \left. -R_2\left(\mathbf{1}-Z_2Z_1^{\ast}\right)^{-1}Z_1^{\ast}
            +R_2\left(\mathbf{1}-Z_2Z_2^{\ast}\right)^{-1}Z_2^{\ast}\right]\left(Z_1^{\ast}-Z_2^{\ast}\right)^{-1}
\nonumber
\end{eqnarray}
with $R_1=\left(Z_1-Z_2\right)^{-1}(Z_1-\mathbf{1})$ and $R_2=\left(Z_1-Z_2\right)^{-1}(Z_2-\mathbf{1})$. 
Summing the four terms in Equation~\ref{eq:GGPI} then results in $Q=E\left\{G^{\top}_{T+1}G_{T+1}\right\}$
from which $\mu_1$ and $\mu_2$ can be extracted. Again, we refer to~\cite{GITHUB} for the details
of the lengthy derivations.
\begin{figure}[tb]
\begin{center}
\includegraphics[width=0.47\textwidth]{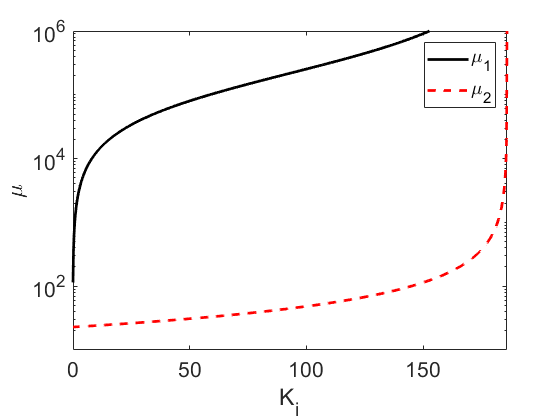}
\includegraphics[width=0.47\textwidth]{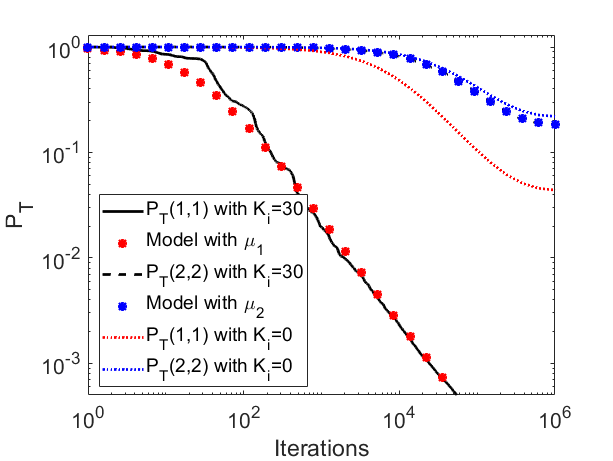}
\end{center}
\caption{\label{fig:Ki}The parameters $\mu_1$ and $\mu_2$ as a function
  of $K_i$ (left) and $P_T(1,1)$ and $P_T(2,2)$ for $K_i=0$ (dotted red and blue) and $K_i=30$
  (solid black) along with the model (red and blue asterisks).}
\end{figure}
\par
The left-hand plot in Figure~\ref{fig:Ki} shows $\mu_1$ and $\mu_2$ as a function of
$K_i$ from zero up to the limit of stability. We observe that increasing $K_i$ always
increases them and thus helps to identify the system parameters. The right-hand
plot shows the evolution of $P_T(1,1)$ and $P_T(2,2)$ for $K_i=0$ (dotted red and blue) 
and  $K_i=30$ (solid black) where $\mu_1\approx 41900$. The faster convergence for the 
larger value of $K_i$ is clearly visible. Moreover, the model values, calculated
from Equation~\ref{eq:pta} with $\mu_1=41900$ and shown as red asterisks, agree well
with those from the simulation. On the other hand, $P_T(2,2)$ for $K_i=0$ and
$K_i=30$ (blue curves) differ only very little. This mimics the behavior of the
proportional controller, given by Equation~\ref{eq:defmu}, where $\mu_1$ is always
larger than $\mu_2$.
\section{Convergence with a PID controller}
\label{sec:appC}
For a PID regulator, the control law for the currents reads
\begin{equation}\label{eq:cpid}
  \vec i_t=K_p\vec v_t -K_d(\vec v_t-\vec v_{t-1}) -K_i\sum_{i=0}^t\vec v_i
\end{equation}
which, upon inserting into Equation~\ref{eq:dsa}, leads to
\begin{equation}
  \vec v_{t+1} = \left(A+B K_p-B K_d\right)\vec v_t +BK_d\vec v_{t-1}- B K_i\sum_{i=0}^t\vec v_i\ .
\end{equation}
As for the PI regulator, we remove the dependence on the sum by taking the difference
\begin{eqnarray}
  \vec v_{t+2}-\vec v_{t+1}
  &=& \left( A+B K_p-B K_d-BK_i\right)\vec v_{t+1}\\
  && \quad- \left(A+B K_p-2BK_d\right)\vec v_{t} - BK_d\vec v_{t-1}\nonumber
\end{eqnarray}
and, after adding $\vec v_{t+1}$ to both sides of the equation and using the Ansatz
$v_{t+1}=Z^t\vec w_1$, we arrive at the characteristic equation
\begin{equation}\label{eq:ZZZ}
  Z^3-\left(\mathbf{1}+A+B K_p-B K_d-BK_i\right)Z^2+ \left(A+B K_p-2BK_d\right) Z +BK_d = 0\ .
\end{equation}
In order to find the roots of this matrix-valued equation, we exploit the fact that all
matrices before the powers of $Z$ can be diagonalized simultaneously with the matrix $V$
\begin{equation}
\mathbf{1}+A+B K_p-B K_d-BK_i=V\Lambda V^{\ast}
\end{equation}
where $\Lambda$ is diagonal with complex-conjugate eigenvalues and $V^{\ast}$ is the hermitian
conjugate of $V$. Using $V$ we bring $A+B K_p-2BK_d$ and $BK_d$  to the same basis
\begin{equation}
  S=V^{\ast}\left(A+B K_p-2BK_d\right) V
  \qquad\mathrm{and}\qquad
  T=V^{\ast}BK_dV
\end{equation}
are all diagonal with complex-conjugate values on the diagonals. This transforms
Equation~\ref{eq:ZZZ} into two complex-valued equations, where one is the complex
conjugate of the other
\begin{equation}
  z^3-\Lambda_{11}z^2+S_{11}z+T_{11}=0
\end{equation}
whose roots $z_1, z_2,$ and $z_3$ can be found numerically, such that we can reassemble
the matrices $Z_i$ with
\begin{equation}
   Z_i=V\left(\begin{array}{cc} z_i & 0 \\ 0 & z_i^{\ast}\end{array}\right) V^{\ast}
\end{equation}
and use them to obtain $\vec y_t$ 
\begin{equation}\label{eq:yPID}
\vec y_{t+1} = V \left[ C_1 Z_1^t +  C_2 Z_2^t +C_3 Z_3^t\right] V^{\ast}\vec w_1
\end{equation}
from which we determine $\vec v_{t+1}=\vec y_{t+1}-\vec y_t$, as for the PI controller in
the previous appendix. We find the constants $C_1$ from matching the boundary conditions
$\vec y_0=\vec y_1=0$ and $\vec y_2=\vec w_1$, by solving the following system of equations
\begin{equation}
  \left(\begin{array}{c} 0\\ 0\\ 1 \end{array}\right) =
  \left(\begin{array}{ccc}
     1 & 1& 1\\ z_1 & z_2 & z_3\\  z_1^2 & z_2^2 & z_3^2
  \end{array}\right)
  \left(\begin{array}{c} c_1\\ c_2\\c_3 \end{array}\right) 
\end{equation}
and assembling the $C_i$ with
\begin{equation}
C_i = V \left(\begin{array}{cc} c_i & 0 \\ 0 & c_i^{\ast} \end{array}\right)  V^{\ast}\ .
\end{equation}
For $\vec v_t$ we obtain
\begin{equation}\label{eq:vPID}
\vec v_t = \left[C_1(Z_1-\mathbf{1})Z_1^{t-1}+C_2(Z_2-\mathbf{1})Z_2^{t-1}+C_3(Z_3-\mathbf{1})Z_3^{t-1}\right]\vec w_1
\end{equation}
Analogously to the procedure in Appendix~\ref{sec:appA} and~\ref{sec:appB}, where we summed over all
earlier noise sources to determine the voltage at time $t$, we express $G_t$ as
\begin{eqnarray}\label{eq:defGPID}
  G_t &=& \left(\begin{array}{rr} (-1+\tR(K_p-K_d))v_r & -v_i \\
                  (-1+\tR(K_p-K_d))v_i& v_r\end{array}\right)_t
       +\tR K_d\left(\begin{array}{lc} v_r & 0 \\ v_i & 0\end{array}\right)_{t-1}
       -\tR K_i\left(\begin{array}{lc} y_r & 0 \\ y_i & 0\end{array}\right)_{t}\nonumber\\
      &=& \vec v_t^{\top} H + \vec v^{\top}_{t-1}J - \vec y^{\top}_{t}K\ .
\end{eqnarray}
$H$ and $J$ are the same as used in Appendix~\ref{sec:appA} and $K$ is defined in
Equation~\ref{eq:defKPI}. 
Definitions for $\bar K$, analogous to those in Equation~\ref{eq:defbJPD}, allow us to
write $G^{\top}_{T+1}G_{T+1}$ as
\begin{eqnarray}
  G^{\top}_{T+1}G_{T+1}
  &=& \bar H \vec v_{T+1}\vec v_{T+1}^{\top} H + \bar H \vec v_{T+1}\vec v_{T}^{\top} J - \bar H\vec y_{T+1}\vec y_{T+1}^{\top} K\nonumber\\
  & & +\bar J \vec v_{T}\vec v_{T+1}^{\top} H + \bar J \vec v_{t}\vec v_{T}^{\top} J - \bar J\vec v_{T}\vec y_{T+1}^{\top} K\\
  & & -\bar K \vec y_{T+1}\vec v_{T+1}^{\top} H - \bar K \vec y_{T+1}\vec v_{T}^{\top} J + \bar K\vec y_{T+1}\vec y_{T+1}^{\top} K\ .\nonumber
\end{eqnarray}
with $\vec y_t$ given by Equation~\ref{eq:yPID} and $\vec v_t$ by~\ref{eq:vPID}, we evaluate all
expectation values and, in this way, determine $Q$ in a similar way done in Equation~\ref{eq:EvvPDapp}.
We refrain from presenting the lengthy calculations to determine $\mu_1$ and $\mu_2$. Instead, we
refer to the code from~\cite{GITHUB} for the details.
\par
We checked the resulting values against simulations and found good agreement similar to that
found on the right-hand side in Figure~\ref{fig:Ki}. Moreover, the limiting cases with
$K_i=0$ agree with the values from Appendix~A and for $K_d=0$ with those from Appendix~B.
\section{Other configurations for system identifications}
\label{sec:appD}
In this appendix we consider complementary definitions of $G_t$ and $\vec y_{t+1}$ to those from
Equation~\ref{eq:defGy}. Let us first assume that the matrix $B$ is constant; this
would imply that $\oo$ is constant or varies very little as a function of the physical 
process one cares about. In that case Equation~\ref{eq:defGy} is replaced by
\begin{equation}
  G_t = \left(\begin{array}{rr} -v_r & -v_i  \\ -v_i & v_r \end{array}\right)_t
  \qquad\mathrm{and}\qquad
  \vec y_{t+1}= \vec v_{t+1}-\vec v_t-B \vec i_t\ .
\end{equation}
The subsequent Equations~21 to~23 remain unaffected, but the remainder of Section~\ref{sec:sysid}
can be simplified, because
\begin{equation}
  G_t^{\top}G_t=(v_r^2+v_i^2)_t{\mathbf 1} = \vec v_t^2 {\mathbf 1}
\end{equation}
is proportional to the unit matrix ${\mathbf 1}$. This renders the fit into two
orthogonal and independent parts; one for each of the fit parameters. To proceed, we
introduce the scalar quantity $p_T$ with $P_T=p_T{\mathbf 1}$ and find that it obeys
\begin{equation}
  p_{T+1}^{-1}=p_T^{-1}+\vec v_T^2\ .
\end{equation}
Here we do not need the Woodbury matrix identity to derive the update for $P_T$ in
Equation~\ref{eq:upp}; simply taking the reciprocal leads to 
\begin{equation}\label{eq:uppD}
  p_{T+1} = \frac{p_T}{1+p_T\vec v_T^2} = \left(1 - \frac{p_T \vec v_T^2}{1+p_T\vec v_T^2}\right)p_T\ ,
\end{equation}
where the second equality proves convenient in the following. This recursion is initialized
with $p_0=1$ in the simulations. As in the main part of this report, we carry $p_0$ through 
all equations, because it carries the inverse units of $\vec v_T^2$. In this simple case,
the update of $\vec q_{T+1}$ in Equation~\ref{eq:upq} becomes
\begin{equation}\label{eq:upqD}
  \vec q_{T+1}=\left[1-\frac{p_T \vec v_T^2}{1+p_T\vec v_T^2}\right]\left( \vec q_T+p_tG_{T+1}^{\top}\vec y_{T+2}\right)\ .
\end{equation}
Equations~\ref{eq:uppD} and~\ref{eq:upqD} only require to calculate the inverse of $1+p_T\vec v_T^2$ and
a few multiplications, rather than the inverse of a $2\times 2$ matrix that is required in 
Section~\ref{sec:sysid}. Note that $P_T=p_T\mathbf{1}$ is diagonal and that makes the two
parameters uncorrelated. Therefore it is possible to determine, for example, the detuning
$q(2)=\Do\dt$ alone by only updating the second component of $\vec q_{T+1}$ in Equation~\ref{eq:upqD}.
\par
Adapting Equations~\ref{eq:uppD} and~\ref{eq:upqD} to time-varying parameters with the help of
the parameter $\alpha=1-1/N_f$ simply involves replacing $p_T$ by $p_T/\alpha$ and leads to 
\begin{eqnarray}
  p_{T+1} &=& \frac{1}{\alpha}\left(1 - \frac{p_T \vec v_T^2}{\alpha+p_T\vec v_T^2}\right)p_T\\
  \vec q_{T+1}&=&\left[1-\frac{p_T \vec v_T^2}{\alpha+p_T\vec v_T^2}\right]
                 \left( \vec q_T+\frac{1}{\alpha}p_tG_{T+1}^{\top}\vec y_{T+2}\right)\ .
\nonumber
\end{eqnarray}
The convergence of the system for different values of $\alpha$ can be done in the same way
as in Sections~\ref{sec:conv} and~\ref{sec:convt} and different controllers as in
Appendices~\ref{sec:appA} to~\ref{sec:appC}.
\par
One might be tempted to treat $B=\oo\dt \tR$ as an independent fit parameter $q(3)$, which
renders Equation~20 to be
\begin{equation}
  G_t = \left(\begin{array}{rrr} -v_r & -v_i & i_r  \\ -v_i & v_r & i_i\end{array}\right)_t
  \qquad\mathrm{and}\qquad
  \vec y_{t+1}= \vec v_{t+1}-\vec v_t\ .
\end{equation}
The remainder of the equations in Section~\ref{sec:sysid} and~\ref{sec:tvp} are
unaffected and remain valid. Running this simulation, however, showed that the
system does not converge to the ``true'' values and that the matrix elements of
the $3\times 3$ matrix $P_T$ stay very large, in particular $P_T(1,1)$ and $P_T(1,3)$.
This indicates the unsurprising fact that $q(1)=\oo\dt$ and $q(3)=\oo\dt\tR$ are
strongly correlated. Independently determining them using only the currents and the
voltages is not possible. 
\end{document}